\documentclass[usenatbib]{mnras}

\usepackage[T1]{fontenc}
\usepackage{ae,aecompl}
\usepackage{outlines}
\usepackage{pythontex}

\usepackage{subfigure}
\usepackage{graphicx}	
\usepackage{amsmath}	
\usepackage{amssymb}	
\usepackage{booktabs}
\usepackage{eso-pic}

\usepackage{ulem}       
\usepackage{url}
\usepackage{times}
\usepackage{array}
\usepackage{dirtytalk}
\usepackage{verbatim}
\usepackage{gensymb}

\usepackage{natbib}

\usepackage{scalerel}
\usepackage{tikz}
\usetikzlibrary{svg.path}

\definecolor{orcidlogocol}{HTML}{A6CE39}
\tikzset{
  orcidlogo/.pic={
    \fill[orcidlogocol] svg{M256,128c0,70.7-57.3,128-128,128C57.3,256,0,198.7,0,128C0,57.3,57.3,0,128,0C198.7,0,256,57.3,256,128z};
    \fill[white] svg{M86.3,186.2H70.9V79.1h15.4v48.4V186.2z}
                 svg{M108.9,79.1h41.6c39.6,0,57,28.3,57,53.6c0,27.5-21.5,53.6-56.8,53.6h-41.8V79.1z M124.3,172.4h24.5c34.9,0,42.9-26.5,42.9-39.7c0-21.5-13.7-39.7-43.7-39.7h-23.7V172.4z}
                 svg{M88.7,56.8c0,5.5-4.5,10.1-10.1,10.1c-5.6,0-10.1-4.6-10.1-10.1c0-5.6,4.5-10.1,10.1-10.1C84.2,46.7,88.7,51.3,88.7,56.8z};
  }
}

\newcommand\orcidicon[1]{\href{https://orcid.org/#1}{\mbox{\scalerel*{
\begin{tikzpicture}[yscale=-1,transform shape]
\pic{orcidlogo};
\end{tikzpicture}
}{|}}}}

\usepackage{hyperref} 

\newcommand{\HEALPIX}{{\textsc{HEALPix}}}

\def \cf2{{\it Cosmicflows-2\,}}
\def \wmap{{\it WMAP\,}}

\def\fnllocal{{f^{\rm local}_{\rm NL}}}

\def\mnras{Monthly Notices of the Royal Astronomical Society}
\def\apj{The Astrophysical Journal}
\def\apjl{Astrophysical Journal Letters}
\def\aap{Astronomy and Astrophysics}

\def\jcap{Journal of Cosmology and Astroparticle Physics}
\def\prd{Physical Review D}

\def\nat{Nature}

\def\apjs{The Astrophysical Journal Supplement Series}

\makeatletter
\let\ftype@table\ftype@figure
\makeatother

\title[Extracting $f_{\rm NL}$ with machine learning]{Constraining primordial non-Gaussianity using Neural Networks}
\author[C.~G.~Nagarajappa \& Y.-Z.~Ma]{Chandan G. Nagarajappa$^{1}$ \orcidicon{0000-0002-8325-7439}, Yin-Zhe Ma$^{2,3,1}$\thanks{Corresponding author: Y.-Z. Ma, \url{mayinzhe@sun.ac.za}}\orcidicon{0000-0001-8108-0986},
\\
$^{1}$  Astrophysics and Cosmology Research Unit, School of Chemistry and Physics, University of KwaZulu-Natal, Westville Campus, \\ Private Bag X54001, Durban, 4000, South Africa \\
$^{2}$ Department of Physics, Stellenbosch University, Matieland 7602, South Africa \\
$^{3}$ National Institute for Theoretical and Computational Sciences (NITheCS), South Africa}

\pubyear{2023}

\begin{document}
\label{firstpage}
\pagerange{\pageref{firstpage}--\pageref{lastpage}}
\maketitle
\begin{abstract}
We present a novel approach to estimate the value of primordial non-Gaussianity ($f_{\rm NL}$) parameter directly from the Cosmic Microwave Background (CMB) maps using a convolutional neural network (CNN). While traditional methods rely on complex statistical techniques, this study proposes a simpler approach that employs a neural network to estimate $f_{\rm NL}$. The neural network model is trained on simulated CMB maps with known $f_{\rm NL}$ in range of $[-50,50]$, and its performance is evaluated using various metrics. The results indicate that the proposed approach can accurately estimate $f_{\rm NL}$ values from CMB maps with a significant reduction in complexity compared to traditional methods. With $500$ validation data, the $f^{\rm output}_{\rm NL}$ against $f^{\rm input}_{\rm NL}$ graph can be fitted as $y=ax+b$, where $a=0.980^{+0.098}_{-0.102}$ and $b=0.277^{+0.098}_{-0.101}$, indicating the unbiasedness of the primordial non-Gaussianity estimation. The results indicate that the CNN technique can be widely applied to other cosmological parameter estimation directly from CMB images.

\end{abstract}
%

\section{Introduction}
\label{sec:intro}
The Cosmic Microwave Background (CMB) radiation is the snapshot of the photon temperature variation across the whole sky at the time of recombination (around 380,000 years after the Big Bang), which characterises crucial information to understand the early Universe~\citep{Scott2010,Dodelson2020}. The CMB temperature and polarization anisotropies provide measurable quantities that can be compared with theories of the early Universe and constrain cosmological parameters~\citep{Komatsu2009,Hinshaw2013,PlanckCollaboration2020}. One of the key parameters of interest is the level of primordial non-Gaussianity in the primordial fluctuations, which is captured by non-linear coupling parameter $f_{\rm NL}$~\citep{Komatsu2001,Planck2015_nG,Planck2018_nG} for the deviation to Bardeen curvature perturbation $\Phi$ ($\Phi_{\rm H}$ in~\citealt{Bardeen1980}), i.e.
\begin{eqnarray}
\Phi(\mathbf{x})=\Phi_{\rm L}(\mathbf{x})+f_{\rm NL}\left(\Phi_{\rm L}^{2}(\mathbf{x}) -\langle \Phi_{\rm L}^{2}(\mathbf{x}) \rangle \right),
\end{eqnarray}
where $\Phi_{\rm L}$ are the Gaussian linear perturbations with zero mean. 

There have been several statistical methods developed to estimate the non-Gaussianity parameter $f_{\rm NL}$ from the CMB maps since the era of the Wilkinson Microwave Anisotropy Probe ({\it WMAP}). One of the most commonly used methods is bispectrum analysis, which measures the correlations between three different Fourier modes ($k_{1},k_{2},k_{3}$) in the CMB temperature and polarization maps and compares with the theoretical models. {\it WMAP} used its five-year data and provided an estimate of the local shape of non-Gaussianity $f^{\rm local}_{\rm NL}$\footnote{The local non-Gaussianity corresponds to the $k$-space configuration as $k_{3}\ll k_{2} \approx k_{1}$, which means the correlations of long-wavelength modes with short-wavelength fluctuations. This $\fnllocal$ is small and undetectable for single-field slow-roll inflation models~\citep{Salopek1990,Falk1993,Gangui1994,Maldacena2003}, but can be large and detectable for multi-field, curvaton scenario~\citep{Lind1997,Lyth2003} or violent, non-linear reheating~\citep{Enqvist2005,Jokinen2006,Chambers2008}.} as $-9 < \fnllocal <111$ at 95\% confidence level (C.L.)~\citep{Komatsu2009}. The ESA's {\it Planck} satellite measured the full-sky CMB fluctuations with higher resolution, and its nominal mission gives $\fnllocal = 2.7 \pm 5.8$~\citep{PlanckCollaboration2014} and full-mission gives $-0.9 \pm 5.1$~($1\sigma$ C.L.,~\citealt{PlanckCollaboration2020,Planck2018_nG}). In addition, the South Pole Telescope (ground-based CMB experiment) has also measured the $f_{\rm NL}$ using the bispectra method, and gives $\fnllocal=420 \pm 350$~\citep{Fergusson2012}. Future surveys such as Simons Observatory~\citep{Ade2019} and CMB Stage-4~\citep{Abazajian2022} will constrain the $\fnllocal$ to higher precision.


In addition to the bispectrum analysis of the CMB, there are several complementary methods to constrain the primordial non-Gaussianity. For example, the Minkowski functionals analysis characterises the topology of the CMB maps using topological descriptors known as Minkowski functionals, and applied the topological estimators to the observed CMB maps~\citep{Hikage2006,Planck2015_nG,Planck2018_nG}. The peak counting method involves counting the number of peaks in the CMB maps and uses this information to constrain $f_{\rm NL}$~\citep{Marian2011}; and the wavelet bispectrum method decomposes the CMB maps into different wavelet scales and uses the bispectrum of each scale to estimate $f_{\rm NL}$~\citep{Curto2011,PlanckCollaboration2014}. Apart from using CMB maps, large-scale structure data has also been used to constrain primordial non-Gaussianity in different ways, which include using Baryon Oscillation Spectroscopic Survey's (BOSS) DR16 data~\citep{Barreira2022,Cabass2022b,Cabass2022a}, using peculiar velocity field scale-dependent bias~\citep{Ma2013} and using 21-cm intensity mapping large-scale correlated bias~\citep{Li2017}.

While the aforementioned methods have been successful in estimating $f_{\rm NL}$ from CMB maps, they can be computationally expensive and time-consuming, and may require careful modelling of the instrumental effects and other sources of uncertainty in the data. In particular, we want to explore the possibility of using neural network method of machine learning to directly recognise the non-Gaussianity pattern in the CMB map. In recent years, machine learning techniques, particularly the neural network, have emerged as a promising technique to deal with large cosmological datasets (see, e.g.~\citealt{George2018},~\citealt{Hezaveh2017},~\citealt{Wang2022},~\citealt{Wang2023}). Neural networks are able to learn complex patterns and correlations in large datasets, and can provide accurate predictions with relatively low computational cost and time. For example, ~\citet{Casaponsa2011} decomposed the \wmap seven-year temperature maps into \HEALPIX\,wavelet and a spherical Mexican hat wavelet, computed the third-order moments of the wavelet coefficients and constrained $\fnllocal$ to be $|\fnllocal| \lesssim 50$. \citet{Novaes2015} used a combined estimator of Minkowski Functionals (MF) and neural networks to constrain $\fnllocal$ to be $\fnllocal = 33 \pm 23$ ($1\sigma$ C.L.) by using {\it Planck} SMICA map. These techniques require the pre-processing the CMB maps into some mathematical basis (e.g. wavelets or MF) to feed the neural networks as the input.


In this work, we adopt a different approach. We will explore the scenario of using neural networks to estimate the primordial non-Gaussianity level from the CMB images directly. We will develop a customized neural network architecture and training process that can handle CMB maps with different levels of primordial non-Gaussianities. We will also test the robustness to the instrumental effects and high-dimensional data representation. 

This paper is organised as follows. In Sec.~\ref{sec: fNL}, we will review the method we use to generate CMB maps with different levels of local shape of primordial non-Gaussianities ($f_{\rm NL}$). In Sec.~\ref{sec:neuralnetwork}, we describe the neural networks, loss function and training process. In Sec.~\ref{sec:results}, we present the results of training and model evaluation. The conclusion is presented in the last section.



\section{Methodology}\label{sec: fNL}
\subsection{Training and Evaluation procedure}
\begin{figure}
    \includegraphics[width=\columnwidth]{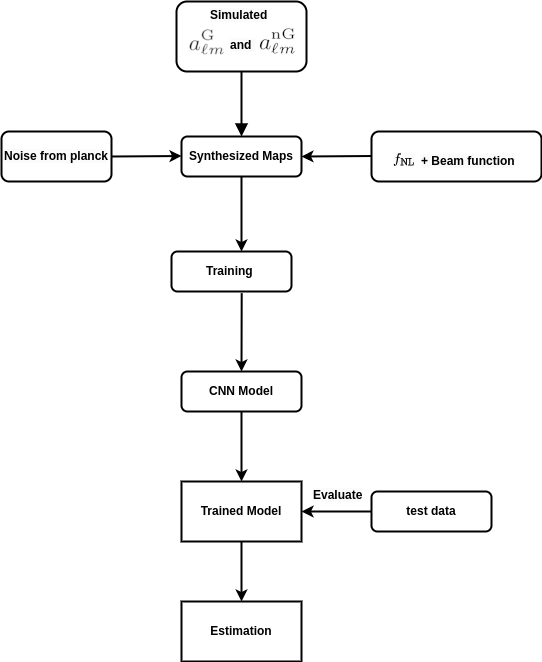}
    \caption{Flowchart depicting the process of generating maps and feeding them into a neural network model, as well as training the model and estimating on test data. The flowchart shows the different stages of the process, including preprocessing, data augmentation, training, and testing, and the different inputs and outputs at each stage. The synthesized maps are the combination of simulated $a^{\rm G}_{\ell m}$ and their corresponding $a^{\rm nG}_{\ell m}$ with varying $f_{\rm NL}$ values, as described by Eq.~(\ref{eq:gng}).}
    \label{fig:Workflow}
\end{figure}

This paper follows a structured flowchart shown in Fig.~\ref{fig:Workflow}. We first simulate the pure Gaussian CMB maps, and add non-Gaussian CMB maps with different $\fnllocal$ values to the Gaussian ones to synthesize the simulated maps. We then convolve them with the {\it Planck} beam function and add {\it Planck} pixel noise. We name them as ``Synthesized Maps''. These synthesized maps form the training data sets used to train the Convolutional Neural Network (CNN) model. Once the network is fully trained, the CNN model is tested on synthetic data to assess its performance. Subsequently, the trained model is employed to estimate $f_{\rm NL}$ values on another set of simulated {\it Planck} maps with different $\fnllocal$ values. This comprehensive approach ensures the model's effectiveness in handling real CMB data and contributes to the understanding of CMB analysis and estimation of $f_{\rm NL}$.

\begin{figure*}
    \centering
    \includegraphics[width=18cm, height=12.5cm]{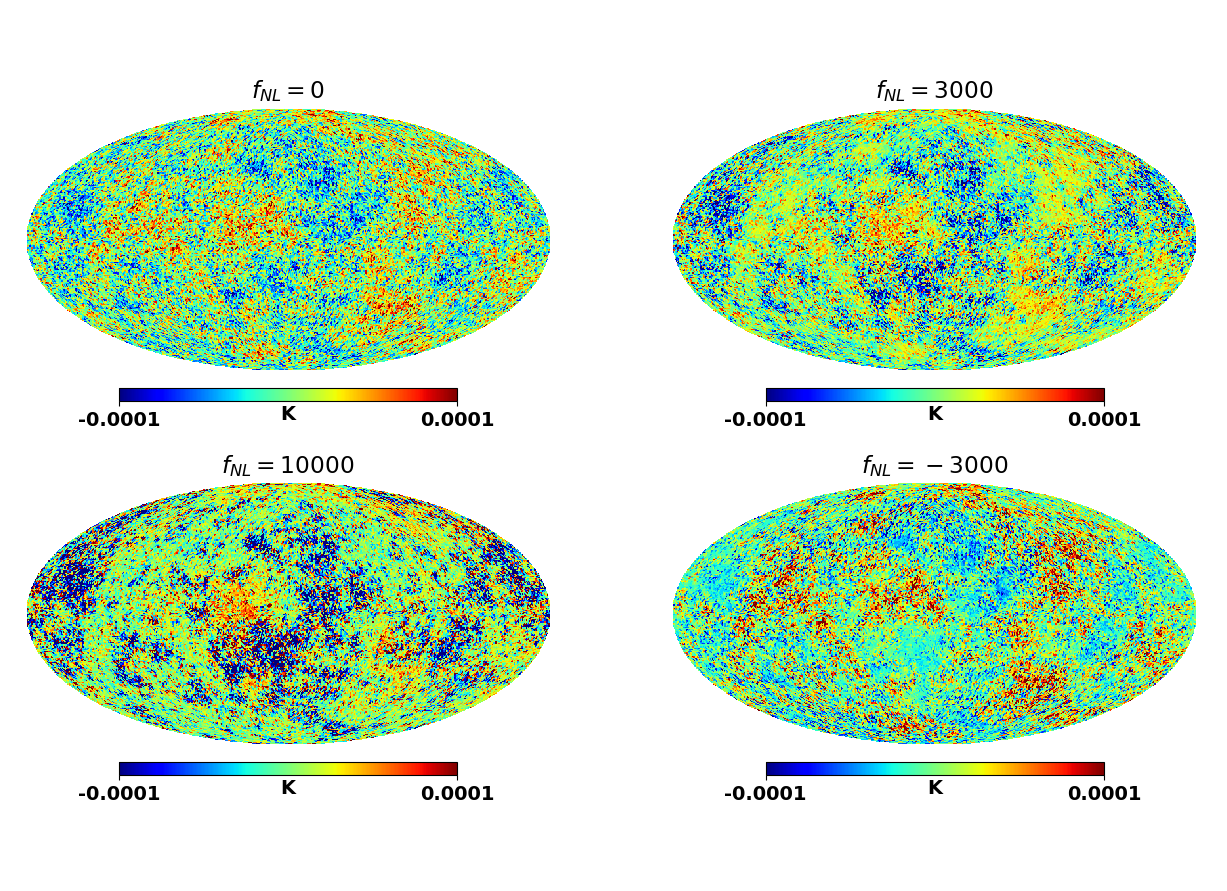}
    \caption{Comparison of CMB maps with different levels of primordial non-Gaussianities (calculated via Eq.~(\ref{eq:sim_alm})). The different panels show the CMB maps with increasing levels of non-Gaussianity. It becomes obvious that above the level of $\fnllocal \gtrsim \mathcal{O}(1000)$ the non-Gaussianity becomes visible in the CMB maps.}
    \label{fig:GnG_maps}
\end{figure*}

\subsection{Simulating CMB Gaussian and non-Gaussian Maps}
We use {\sc camb} to calculate the temperature angular power spectrum of the CMB, by using the {\it Wilkinson Microwave Anisotropy Probe} ({\it WMAP}) five-year data with the BAO and Type-Ia supernovae best-fitting parameter values: $\Omega_\Lambda = 0.721$, $\Omega_{\rm c} h^{2} = 0.1143$, $\Omega_{\rm b} h^{2} = 0.02256$, $h = 0.701$, $n_{\rm s} = 0.96$, $\tau = 0.084$ and $\Delta^{2}_{\mathcal{R}}\,(k_{\ast}=0.002\,{\rm Mpc}^{-1}) =2.457 \times 10^{-9}$~\citep{Komatsu2009,Dunkley2009}. The reason we use this set of parameters is that we will utilise the non-Gaussian maps generated from the ``Optimized Quadrature Scheme'' by~\citet{Elsner2009}, which takes the values in the above set. We then simulate $1000$ pure CMB maps from the outputted theoretical angular power spectrum of CMB ($C_{\ell}$) out to $\ell_{\rm max}=1024$. For the non-Gaussian component of the map ($a^{\rm nG}_{\ell m}$), we utilise the Elsner \& Wandelt ``Optimized Quadrature Scheme'' non-Gaussian maps~\citep{Elsner2009}, and download $1000$ of them from their website\footnote{\url{http://planck.mpa-garching.mpg.de/cmb/fnl-simulations/}}.

We then combine them to compute the synthesized maps with different values of $\fnllocal$:
\begin{eqnarray}
a_{\ell m} = a^{\rm G}_{\ell m} + \fnllocal a^{\rm nG}_{\ell m}.
\label{eq:gng}
\end{eqnarray} 
In Fig.~\ref{fig:GnG_maps}, we show the Gaussian and non-Gaussian maps with different levels of $\fnllocal$ values. One see clearly that above the level of $\fnllocal \gtrsim \mathcal{O}(10^{3})$ the non-Gaussianity of the map becomes visible. But so far, $a_{\ell m}$ in Eq.~(\ref{eq:gng}) is a theoretical simulation which does not include observational effects. We further convolve the $a_{\ell m}$ in Eq.~(\ref{eq:gng}) with Gaussian beam function and add the {\it Planck} instrumental noise to mimic the observational map.

For beam convolution, we just simply multiply Eq.~(\ref{eq:gng}) with a Gaussian kernel
\begin{eqnarray}
a_{\ell m} \rightarrow a_{\ell m}e^{-\ell^{2}\sigma_{\rm b}^{2}/2},
\end{eqnarray}
where $\sigma_{\rm b}=\theta_{\rm FWHM}/\sqrt{8\ln 2}$, and $\theta_{\rm FWHM}=5\,{\rm arcmin}$ is the {\it Planck} Full-Width-Half-Maximum beam size.

\subsection{Noise}
\begin{figure*}
    \includegraphics[width=19cm, height=17.cm]{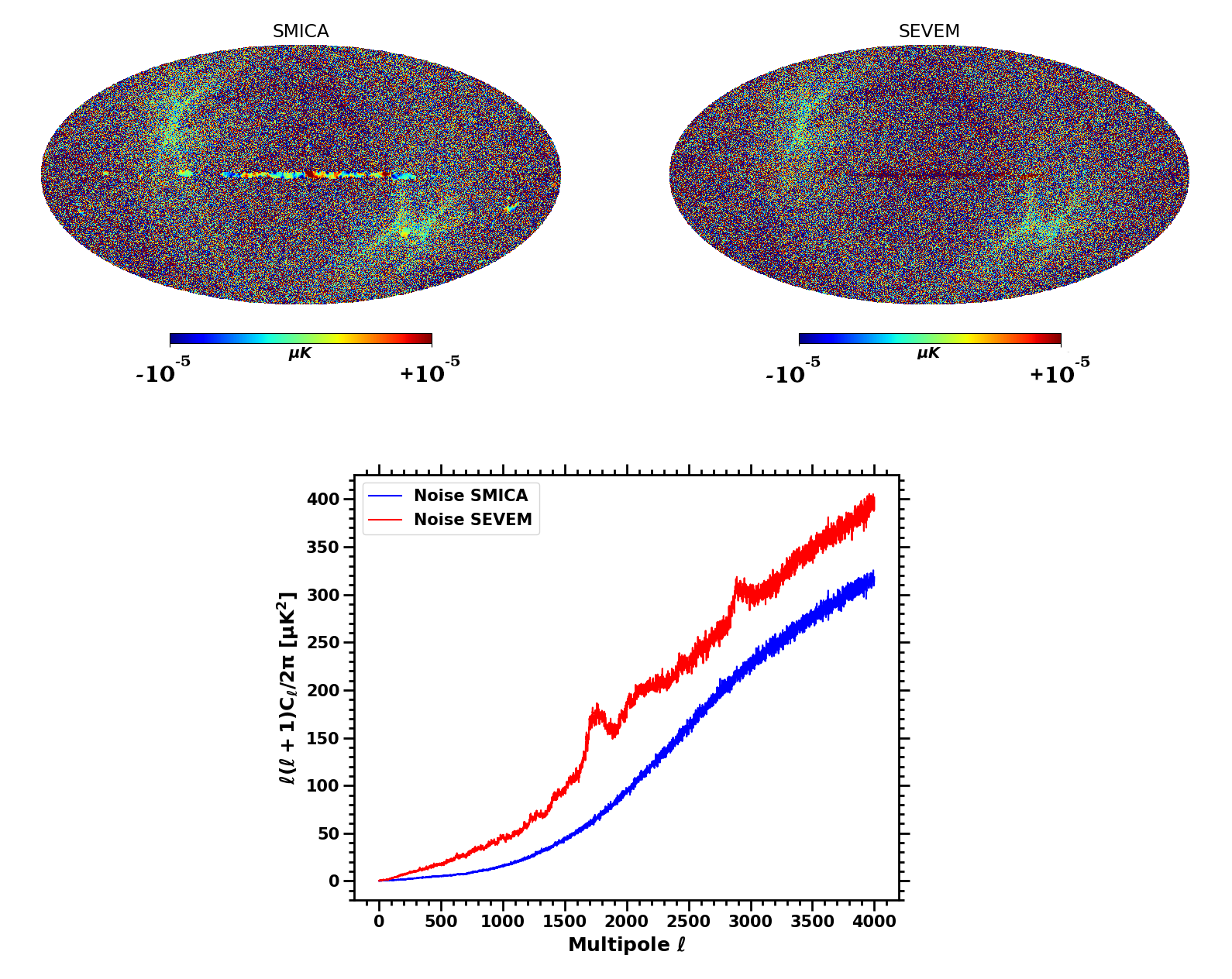}
    \caption{{\it Top Left} and {\it Top Right} panels are the noise maps of {\it Planck} SMICA and SEVEM CMB maps, which are obtained by using the ``Half-Ring Half-Difference (HRHD)'' method. {\it Bottom} panel compares the noise power spectra of SMICA and SEVEM maps.}
    \label{fig:noise}
\end{figure*}

We now simulate the noise of the CMB map. To mimic the true {\it Planck} mission data, we download the {\it Planck} maps and use the ``Half-Ring Half-Difference (HRHD)'' method to estimate the noise power spectrum. Basically, we calculate 
\begin{eqnarray}
{\rm HD} = \frac{\delta_1 - \delta_2}{2},
\label{eq:HRHD}
\end{eqnarray}
where $\delta_{1}$ and $\delta_{2}$ are the splits of the data from {\it Planck} detectors into two halves. Therefore the difference in Eq.~(\ref{eq:HRHD}) makes the signal cancel out but keeps the noise fluctuations. We do this practice for both {\it Planck} SMICA and SEVEM maps, and the resultant HD maps are shown in the upper panels of Fig.~\ref{fig:noise}. One can see that the noise across the whole sky is almost homogeneous except for the ecliptic pole region, which is due to more scannings on these two regions.

\begin{figure}
    \includegraphics[width=\columnwidth]{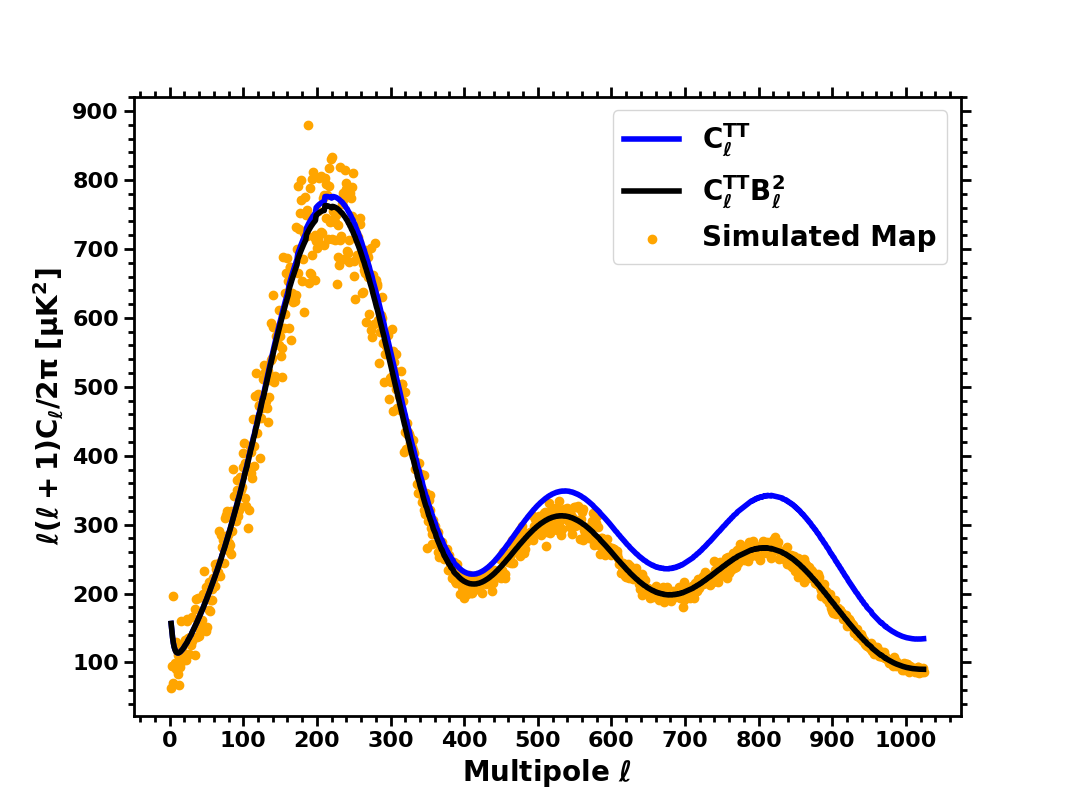} 
    \caption{Comparison between theoretical CMB power spectrum $C^{\rm TT}_{\ell}$, convolved spectrum $C^{\rm TT}_{\ell}B^{2}_{\ell}$ and the estimated power spectrum from simulated maps.}
    \label{fig:powerspectra}
\end{figure}

We then calculate the noise power spectrum ($N_{\ell}$) from the HD maps, and the results are shown in the lower panel of Fig.~\ref{fig:noise}. One can see that the noise power spectrum of SMICA is slightly higher than SEVEM map, which might be due to the difference in the component separation methods. We use the resultant noise power spectrum $N_{\ell}$ to simulate the noise map ($n_{\ell m}$) and add to the convolved CMB map to resemble the total (simulated) CMB map
\begin{eqnarray}
a^{\rm tot}_{\ell m}=a_{\ell m}e^{-\ell^{2}\sigma^{2}_{\rm b}/2}+n_{\ell m}, \label{eq:sim_alm}
\end{eqnarray}
which ensures the training set is as close to the true {\it Planck} map as possible. In Fig.~\ref{fig:powerspectra}, we show the comparison between theoretical CMB power spectrum $C^{\rm TT}_{\ell}$ (blue solid line), convolved spectrum $C^{\rm TT}_{\ell}B^{2}_{\ell}$ (black solid line) and the estimated power spectrum from simulated maps (yellow dots). One can see that the power spectrum from the simulated map is very close to the convolved CMB power spectrum (black line), which is because on scales of $\ell \leq 1000$, the noise level is not very high compared to the intrinsic CMB level. We nonetheless include the noise simulation for the accuracy of the study.

\section{Convolutional Neural Network}
\label{sec:neuralnetwork}
\subsection{Single Neuron} 
\begin{figure}
    \includegraphics[width=\columnwidth]{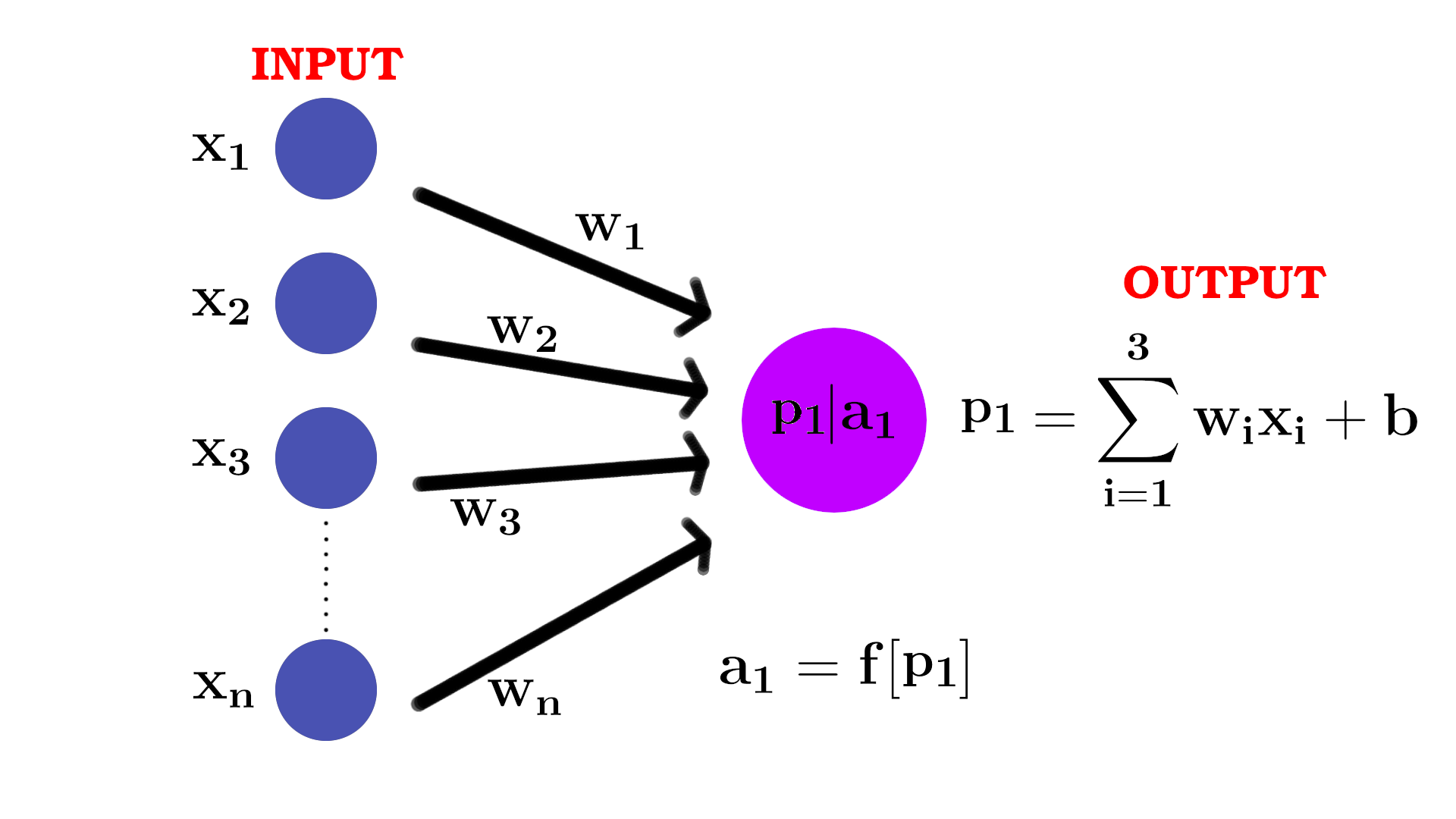}
    \caption{A single neuron with an input and its corresponding output. The input is multiplied by a weight, which is then added to a bias term. The resulting value is passed through an activation function, which produces the output of the neuron.}
    \label{fig:Single Neuron ML}
\end{figure}

A typical single neuron used in machine learning is a fundamental building block of artificial neural networks. It represents the basic computational unit that processes inputs and produces an output as shown in Fig.~\ref{fig:Single Neuron ML}. The neuron receives inputs from various sources, each multiplied by corresponding weights. These inputs are then summed together, including a bias term. The weighted sum is passed through an activation function to introduce non-linearity and determine the neuron's output.

The mathematical representation of a single neuron in a neural network is given by,
\begin{equation}
    p_i = \sum_{i=1}^{n} w_i x_i + b . 
\end{equation}
The neuron takes input signals $(x_1, x_2, ..., x_n)$, and computes a weighted sum of these inputs, denoted by the symbol $z$. The weights used in the summation are denoted by $(w_1, w_2, ..., w_n)$, and the bias term is denoted by $b$. 

The input signals $(x_1, x_2, ..., x_n)$ can represent any number of features or attributes that describe the input data, such as pixel values in an image. The weights $(w_1, w_2, ..., w_n)$ determine the strength of the connection between each input signal and the neuron. Alternatively, one can think of the weights as representing the importance of each input feature for the task at hand. 

The bias term, $b$, is an additional parameter that is added to the weighted sum. It can be regarded as the representation of the neuron's willingness to fire even when all input signals are zero. In this sense, the bias term provides a measure of the neuron's overall activity level. 

Once the weighted sum $p$ is computed, it is typically passed through an activation function, such as the sigmoid function or ReLU function, to introduce non-linearity into the model and allow the neuron to model complex mappings between the input and output.

\subsection{Neural Network}
\begin{figure*}
    \includegraphics{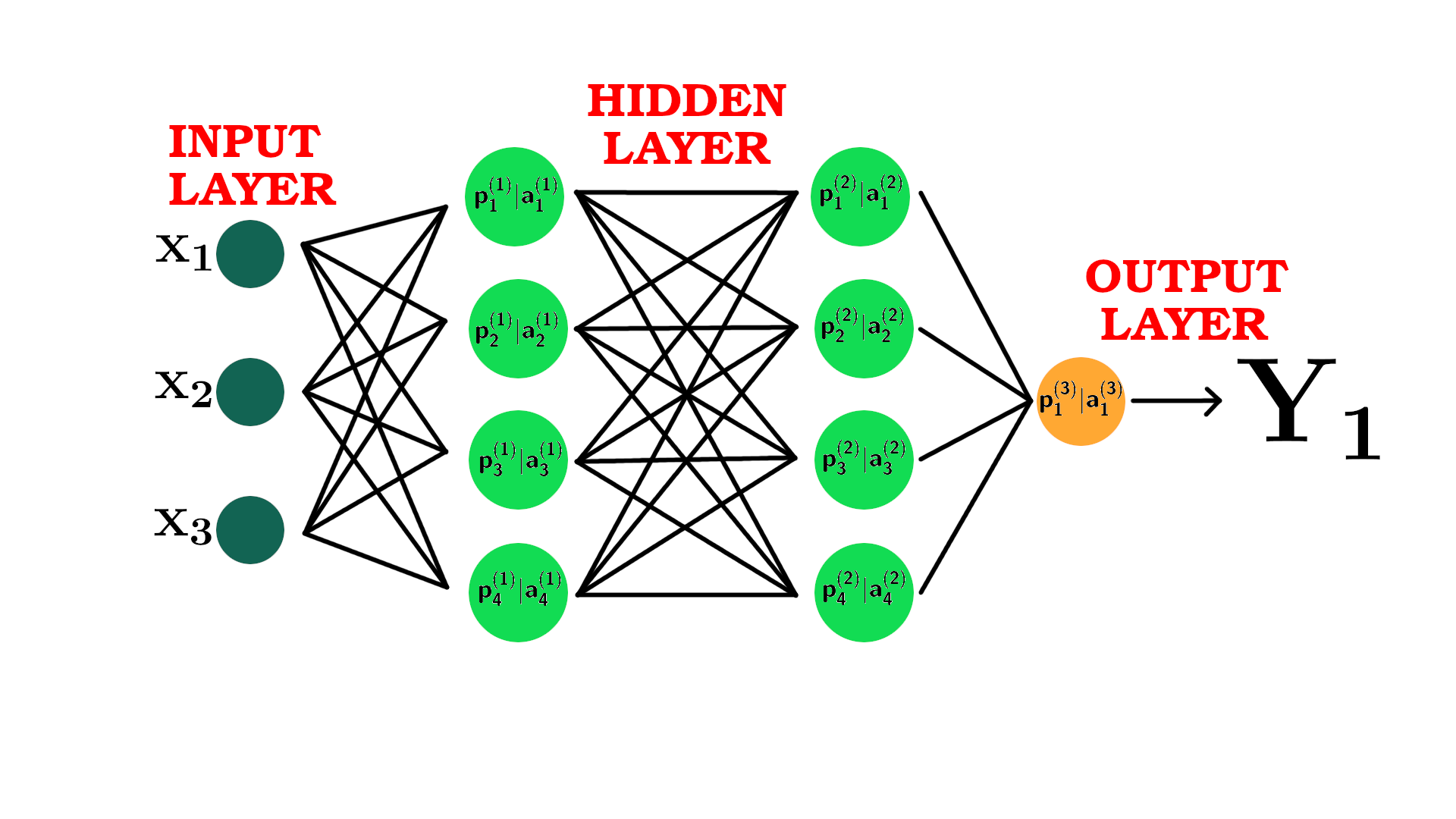}
    \caption{The image depicts a neural network with an input layer, multiple hidden layers, and an output layer. In the above neural network, each circle in a hidden layer represents a neuron, typically displayed in light green. These neurons are interconnected by lines, which signify the connections between them. The output of one neuron can be passed to the inputs of neighbouring neurons, forming the network structure. The yellow circle denotes the output layer, which may consist of a single neuron or multiple neurons depending on the specific task the network is designed to perform. Finally, the output layer produces the final result, which could be a classification or prediction based on the input data. The image demonstrates the complexity of neural networks and how they can be used for tasks such as image recognition, natural language processing, and more.}
    \label{fig:simple CNN structure}
\end{figure*}

A typical neural network is composed of multiple layers of interconnected neurons depicted in Fig.~\ref{fig:Single Neuron ML}, where each neuron performs a specific computation. These neurons are organized into different layers, including input layer, hidden layers, and output layer. A typical neural network is shown in the Fig.~\ref{fig:simple CNN structure}.
 
The input layer of a neural network receives the raw input data (deep green dots on the left side of Fig.~\ref{fig:simple CNN structure}). Each neuron in the input layer represents a specific feature or attribute of the input data. The number of neurons in a neural network is directly related to the dimensionality of the input data. In neural networks, the dimensionality of input data represents the number of variables or attributes describing each data point. In the case of our data with a shape of $(256, 192, 3)$, each data point comprises multiple values due to the presence of three channels. Consequently, the dimensionality is calculated as the total number of points in the 2D array $(256 \times 192)$ multiplied by the number of channels (3). This yields a threefold increase in dimensionality compared to traditional methods that utilize single-channel convolutional neural networks (CNNs). The increase in the number of channels results in a higher dimensional input, which means the model needs to process more data at once. This can lead to increased computational requirements and potentially longer training times. However, the benefits of having more information in the form of additional channels outweighs in our case, as the model can learn more complex patterns and representations from the additional data.

Hidden layers are intermediate layers between the input and output layers (shallow green dots in the middle of Fig.~\ref{fig:simple CNN structure}). They are responsible for learning and representing complex patterns and relationships within the data. Each neuron in the hidden layers receives inputs from the previous layer, applies a weighted sum of these inputs, and passes the result through an activation function. The number of hidden layers and the number of neurons in each hidden layer can vary depending on the complexity of the problem and the desired model capacity.

The output layer represents the final prediction or output of the neural network (yellow dots on the right side of Fig.~\ref{fig:simple CNN structure}). It consists of neurons that produce the desired output based on the learned representations from the hidden layers. The number of neurons in the output layer depends on the specific task. In our work, the number of output neurons corresponds to the single output neuron, because it is a regression task.

\subsection{Convolutional Neural Network (CNN)}
For our regression work, we use a CNN model to carry out the regression task. Our input data is a 2D array of CMB temperature maps. The model uses the Keras deep learning library's Sequential API, which is designed to make it easy to create neural networks consisting of multiple layers. This specific neural network model includes convolutional layers, pooling layers, and fully connected layers, which are typical for image-related problems.

A CNN is a deep learning architecture composed of multiple layers of interconnected neurons, including the typical single neurons described earlier. A CNN is a specialized type of neural network that is specifically designed for processing structured grid-like data, such as images or sequences and has become highly effective in various computer vision tasks.

CNNs utilize specialized layers, namely convolutional layers and pooling layers, which are not typically found in traditional neural networks. Convolutional layers apply convolutional operations to the input data, using learnable filters or kernels to extract local features. These filters slide over the input, computing element-wise multiplications and summations. Pooling layers then downsample the feature maps obtained from the convolutional layers, reducing their spatial dimensions while preserving important features. These layers enable CNNs to capture spatial hierarchies and translation-invariant representations. The earlier layers of a CNN learn low-level features, such as edges or corners, while deeper layers learn higher-level features, such as textures or object shapes. This feature learning capability allows CNNs to effectively extract and represent complex patterns in the CMB data, making them well-suited for our regression task. The following Fig.~\ref{fig:Convolutional Neural Network (CNN) architecture} shows a general CNN structure.

\begin{figure*}
    \centering
    \includegraphics[scale=0.45]{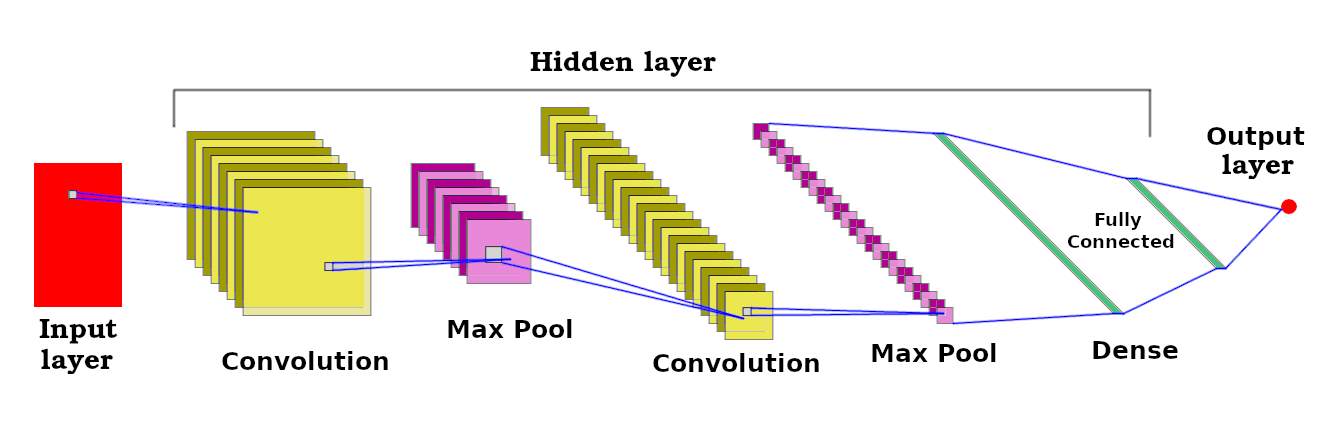}
    \caption{The diagram of a convolutional neural network (CNN) structure. It includes input data, convolutional layers, activation functions, pooling layers, fully connected layers, and an output layer. The input data is passed through the convolutional layers, which extract features from the input data. Activation functions introduce non-linearity to the network, and pooling layers downsample the output of the convolutional layers. The fully connected layers take the output of the pooling layers and produce a final output. The output layer is responsible for producing the final prediction based on the input data.}
    \label{fig:Convolutional Neural Network (CNN) architecture}
\end{figure*}

Our CNN model used for regression consists of 10 layers including the input layer, convolutional layers, dense layers, and the output layer. The breakdown of the layers are:

\begin{itemize}
\item Input Layer: The model expects input images with a shape of $(256, 192, 3)$. This means the input array must have a height of $256$, width of $192$, and three channels. The input layer with appropriate shape is represented in rectangular red box in Fig.~\ref{fig:Convolutional Neural Network (CNN) architecture}.

\item Convolutional Layers:

\begin{enumerate}
    \item  The first convolutional layer has 32 filters of size $(3,3)$ and uses the ReLU activation function. It takes the input image shape. The convolutional layer is part a of hidden layer which is represented in Fig. ~\ref{fig:Convolutional Neural Network (CNN) architecture} in yelow rectangular boxes.
    \item A max-pooling layer with a pool size of $(2,2)$ follows each convolutional layer. It is also a part of a hidden layer. This layer downsamples the feature maps obtained from the previous layer. A max pool layer is represented in pink rectangular box immediately after convolutional layer represented in Fig. ~\ref{fig:Convolutional Neural Network (CNN) architecture}.  It is possible to have a convolutional neural network (CNN) without max pooling layers. While max pooling is a common technique used in CNNs to reduce the spatial dimensions of feature maps and prevent overfitting, it is not an essential component of a CNN.
\end{enumerate}

\item Fully Connected Layers:

\begin{enumerate}
\item After the convolutional layers, a flatten layer is added to convert the 2D feature maps into a 1D vector.
\item Then, a series of dense also known as fully connected layer (green shaded region in Fig. ~\ref{fig:Convolutional Neural Network (CNN) architecture}) layers follow, with $256$, $128$, and $64$ neurons, respectively. Each dense layer uses the ReLU activation function to introduce non-linearity and capture complex relationships in the data.
\item The last dense layer has a single neuron with a linear activation function, which makes it suitable for regression tasks.
\end{enumerate}

\item Model Output: The final layer of the model (represented as red dot in the Fig. ~\ref{fig:Convolutional Neural Network (CNN) architecture}) consists of a single neuron with a linear activation function. This indicates that the model is designed for regression tasks, where it aims to predict a continuous numerical value.

The given CNN model has approximately $1,479,721$ trainable parameters for the input of shape $(256,192,3)$. The number $3$ in $(256,192,3)$ refers to the channel which is a feature related to CNNs. During the convolution operation, the filters (also known as kernels) are applied to the input data slide over the data, computing dot products between the filter and the corresponding input values. The output of this operation is called the feature map, which represents a set of learned features or patterns in the input data. Each channel applies different filters. Usually the input data will be of one-channel meaning $(256,192,1)$ (also written as $(256,192)$) but we stack the array to make it suitable for three-channel CNN so more filters are applied to extract features along the multiple channels.

When the model is compiled using the method, we specify the loss function, optimizer, and evaluation metrics we will use for the training. The optimization algorithm sends error signals, usually in the form of gradients, back through the network to fine-tune the weights of the nodes in the network, which is the essence of backpropagation. The compiling method sets up the backpropagation algorithm to be used during the training process.

Let $L$ be the loss function, $p$ be the output of a particular layer, $x$ be the input to that layer, $w$ be the weights of that layer, $b$ be the bias of that layer, and $\delta$ be the error signal at that layer. The back propagation algorithm can be summarized as follows:

\begin{enumerate}

\item First, compute the error signal $\delta$ for the output layer,

\begin{equation}
    \delta = \frac{\partial x}{\partial p}     .
\end{equation}

\item Then, compute the error signal $\delta$ for the previous layer using the current layer's error signal
\begin{equation}
    \tilde{\delta} = w^T \cdot \delta \cdot g'(x)  ,
\end{equation}
where $T$ is the transpose operator, and $g'(x)$ is the derivative of the activation function used in the previous layer.

\item The gradient of the loss function with respect to the weights and biases of the current layer are
\begin{eqnarray}
\frac{\partial L}{\partial w} &=& \delta \cdot x \nonumber \\   .
\frac{\partial L}{\partial b} &=& \delta   .  
\end{eqnarray}

\item Then, update the weights and biases of the current layer using an optimization algorithm, such as stochastic gradient descent:
\begin{eqnarray}
w_{\rm new} &=& w - \eta \cdot \frac{\partial L}{\partial w}  \nonumber \\
b_{\rm new} &=& b - \eta \cdot \frac{\partial L}{\partial b} ,
\end{eqnarray}
where $\eta$ is the learning rate, which is a hyperparameter that determines the step size of the optimizer while updating the weights of the neural network during training. Therefore, $\eta$ is a critical parameter that needs to be tuned to achieve the best performance of the model. $\eta$ is tuned by trial and error method, or grid search method, or by adopting decaying learning rate technique. In our case, the performance is achieved with a constant learning rate of $0.001$, and we arrived at the value by trial and error method and further tuning during the runs was not necessary once the model was initialized.
    
\end{enumerate}

\end{itemize}

The above steps are repeated for all previous layers in the neural network. The above equations are then used iteratively for multiple epochs until the loss function converges and the neural network is trained. The back propagation algorithm is used in our CNN model to adjust the weights of the neural network to minimize the specified loss function and improve the model's performance on the evaluation metrics.


\subsection{Loss function}
In machine learning, a loss function is used to measure the difference between the predicted outputs of a model and the actual values. The goal of the model is to minimize this loss function during training, which improves the accuracy of predictions.

The choice of loss function depends on the type of problem being solved. For regression problems, where the output is a continuous value, the mean squared error (MSE), the mean absolute error (MAE), or the Huber loss is commonly used. 

Below are the two commonly used loss function
\begin{itemize}
    \item Mean Square Error (MSE).
    \begin{eqnarray}
    l(y_i,y_o) = \frac{1}{n}\left(\sum(y_i - y_o)\right)^2 . 
\end{eqnarray}

\item Mean Absolute Error (MAE)
\begin{eqnarray}
    l(y_i,y_o) = \frac{1}{n}\sum\left| y_i - y_o \right| . 
\end{eqnarray}

\item Huber Loss
\begin{eqnarray}
    l(y_i,y_o) = 
\begin{cases}
    \frac{1}{2} (y_i,y_o)^2 ,& \text{for} |(y_i,y_o)|\leq \delta\\
    \delta \cdot \left(|y_i,y_o|-\frac{1}{2} \delta \right),              & \text{otherwise}.
\end{cases}
\end{eqnarray}
The Huber Loss function effectively combines two components for handling errors differently, with the transition point between these components determined by the threshold $\delta = (y_i,y_o)$.
\end{itemize}

These loss functions are used during training to optimize the parameters of the neural network in order to minimize the difference between the predicted and actual outputs.

\subsection{Data Preparation}
\subsubsection{$z$-scaling}

After obtaining the simulated $a_{\ell m}$ maps through Eq.~(\ref{eq:sim_alm}), we convert the $a_{\ell m}$s into a full-sky map with $n_{\rm side}=64$, resulting in a float array of size 49,152. The histogram of the array values is very close to Gaussian distribution with values in order of $\mathcal{O}(10^{-5})$. Because of the small values in the array which might be insensitive to CNN recognition, we employ the $z$-scaling, which preserves the Gaussian shape of the array while amplifying the value of each data point. This ensures that the CNN can effectively recognize and learn from the changes in the values, even though they are initially very small. As a result, the $z$-scaling technique allows for better sensitivity and interpretation of the data by the CNN.


\begin{figure*}
    \includegraphics[width=19cm]{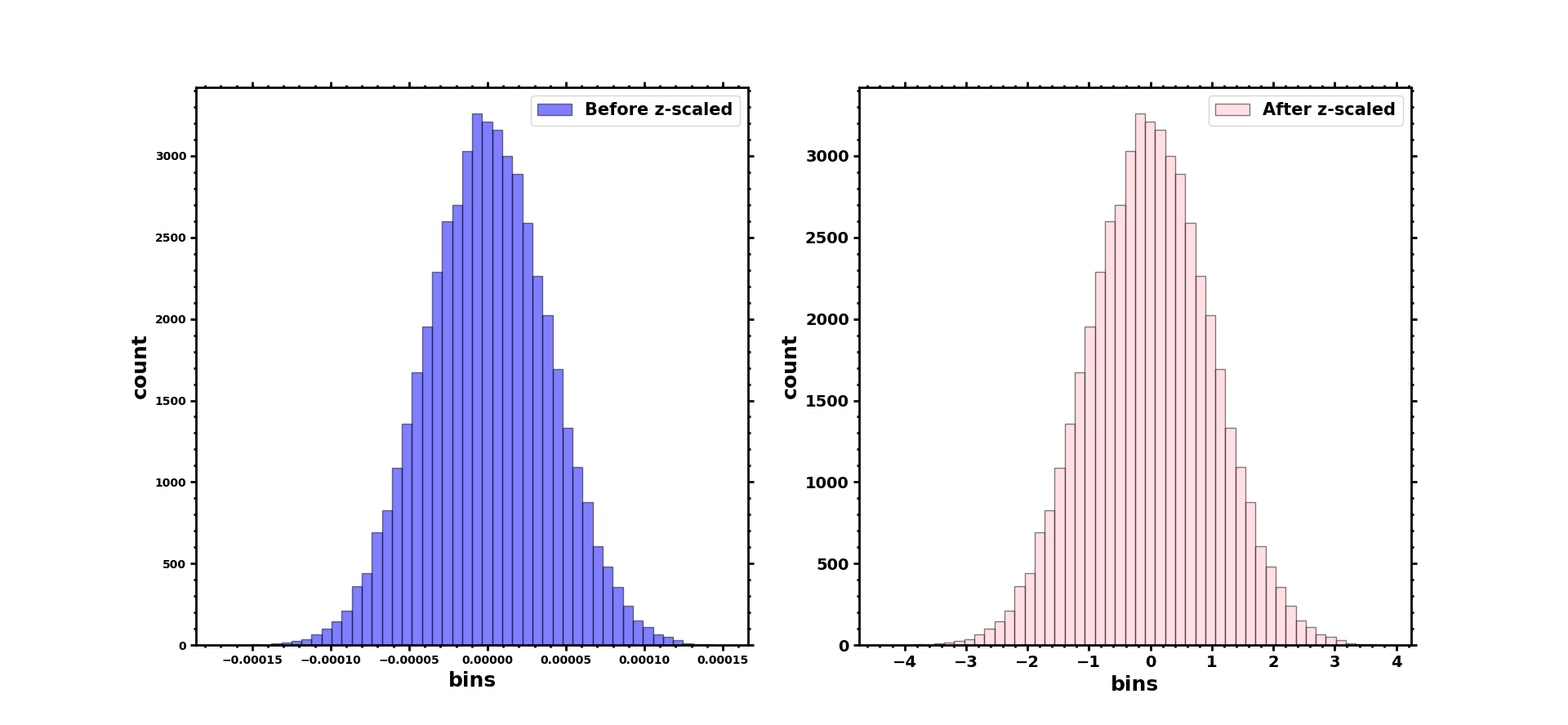}
    \caption{The histogram of the array values before and after $z$-scaling are shown in blue and light pink, respectively. The histogram in light pink shows the distribution of array values after applying $z$-scaling with the mean and standard deviation corresponding to each map (Eq.~(\ref{eq:z_scaling})). The $x$-axis represents the bin range, while the $y$-axis shows the frequency of values in that bin. The sample plot reveals the spread and shape of the data distribution for each map array. After $z$-scaling, the plot shows a roughly symmetric distribution with a peak around zero, indicating that the mean has been shifted to zero. This type of preprocessing is often applied to input data in machine learning models to improve training performance and accuracy.}
    \label{fig:Sample array z-scaled}
\end{figure*}

We apply the $z$-scaling as follows:
\begin{eqnarray}
    z_i = \frac{x_i - \mu}{\sigma}, \label{eq:z_scaling}
\end{eqnarray}
$z_i$ is the standardized value ($z$-score) of the ith data point, $x_i$ is the original value of the data point, $\mu$ is the mean of the data distribution in the given array which is $z$-scaled, and $\sigma$ is the standard deviation of the data distribution in the given array. The mean value of the array for one of the sample map before scaling was $4.822 \times 10^{-8}$ and the standard deviation was $3.91 \times 10^{-5}$. After applying $z$-scaling, the mean value of the array shown in the Fig. ~\ref{fig:Sample array z-scaled} has been shifted very close to zero, $\mu = -2.02 \times 10^{-17} $ and the standard deviation has been scaled to $\sigma = 1$. After $z$-scaling the values are scaled approximately between -4 to 4 for the illustrated map in Fig. \ref{fig:Sample array z-scaled}, making it convenient for the CNN to apply filters to extract features. This transformation preserves the shape of the distribution, meaning that the relative positions and spread of the data points remain unchanged, but they are now measured in standard deviation units instead of their original units. 

This process helps to amplify the values and make them more discernible to the CNN, while maintaining the Gaussian distribution of the data. The $z$-scaling is applied to the each map in the training dataset separately. This approach ensures that the array values are more sensitive to the CNN model. By applying the scaling independently to each map, we prevent any data leakage between different maps, which could potentially compromise the performance of the model on the unseen data. By treating each map independently during the scaling process, we maintain the integrity of the training dataset and ensure that the CNN learns the patterns and variations specific to each map. This approach helps the model generalize better to the new and unseen dataset, as it has not been exposed to any information from other maps.

The histogram in the Fig.~\ref{fig:Sample array z-scaled} illustrates the distribution of values in the array. This scaling process ensures that the CNN can effectively capture the subtle variations and patterns within the data, enabling better interpretation and learning. 

\subsubsection{2D Conversion}
Because the power of the CNN lies in its ability to capture shapes, features, and depths in images, to leverage this capability, we utilize the {\sc HEALPix} module to project the one-dimensional array into a two-dimensional shape.

\begin{figure}
    \centering
    \includegraphics[width=0.5\linewidth]{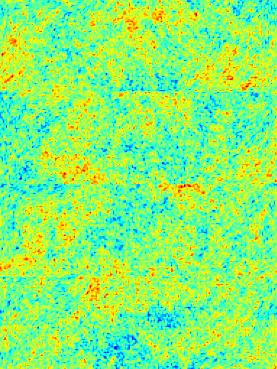}
    \caption{One of the sample map after 2D conversion from the Mollweide projection to the rectangular array of the shape $(256, 192)$. In \HEALPIX, the celestial sphere is divided into a hierarchical, non-overlapping arrangement of equal-area pixels. These pixels are organized with multiples of $12$, ensuring consistency across the dataset~\citep{Healpix2005}. To convert a Mollweide projection to a 2D rectangular array, such as a shape of $(256, 192)$ shown in this figure, the dimensions are calculated as $3\times n_{\rm side}$ in one direction, and $4\times n_{\rm side}$ in the other direction (i.e. $3n_{\rm side} \times 4n_{\rm side}$ in total). This 2D conversion is suitable for machine learning algorithms as long as the consistency in the map division is maintained throughout the dataset.}
    \label{fig:2dmap}
\end{figure}

The array after $z$-scaling is still a 1D array, which is then converted into 2D map. For a map of $n_{\rm side} = 64$, the array has $49,152$ elements. Following \citet{Wang2022b}, we transform the effective map into a two-dimensional array of size (256, 192) using the code {\sc cmbNNCS}\footnote{https://github.com/Guo-Jian-Wang/cmbnncs}, where the values from the $z$-scaled array are preserved without any clipping or truncation of data points. This approach ensures that no information is lost compared to conventional methods that convert the map into a Mollweide projection or a PNG image with a white background. Such conventional methods can introduce systematic errors and distortions to the data, compromising its accuracy. 

By directly converting the map into a two-dimensional array as show in the Fig. \ref{fig:2dmap} while preserving the $z$-scaled values, we maintain the integrity of the data, allowing the CNN to effectively learn and extract meaningful features for improved analysis and prediction. This process gives us the map of size $(256, 192)$.

\subsubsection{Array stacking}
The 2D conversion of the map results in an array of size $(256, 192)$, which is suitable for a CNN to learn features and shapes within the data. However, it does not capture the depth information. In Fig.~\ref{fig:GnG_maps}, we can observe that different values of $f_{\rm NL}$ uniformly change the values of the array, but the underlying features remain consistent. The features of the map are altered when different realizations of the simulated maps are combined linearly. This would have been good to estimate $f_{\rm NL}$ within the same realization because features will not change for the same realization of the CMB map, but this does not suit our application because we have different realisations of CMB.

Through array stacking, we transform each $(256,192)$ map from a single-channel format, suitable only for a single-channel CNN, into a $(256,192,3)$ format, enabling compatibility with a three-channel CNN. This multi-channel transformation empowers the CNN to extract intricate features from the CMB map, enhancing its ability to capture nuanced details and varying intensities across the data depth. By leveraging information from all three channels, the CNN's predictive capabilities are strengthened, resulting in improved prediction accuracy.

During training, a dataset comprising 9500 maps is utilized, shaping the input data into a data cube of shape $(9500, 256, 192, 3)$ to adhere to the conventions of {\tt numpy} arrays and meet the requirements of CNNs. The `three' designation signifies the compatibility with CNNs operating on three-channel inputs.

\subsection{Training}
The dataset of 10,000 maps was generated by combining Gaussian and non-Gaussian maps provided by~\citet{Elsner2009}, which represent different realisations. We then multiply a range of $\fnllocal$ values in between $-50$ to $50$ to the $a^{\rm nG}_{\ell m}$ maps and add them to the Gaussian CMB maps (Eq.~(\ref{eq:gng})).

During the training process, the model's parameters are adjusted to minimize the difference between the predicted and actual values of the $\fnllocal$ maps in the training dataset. This is done using a technique called backpropagation, which is a type of gradient descent algorithm. The validation dataset is used to monitor the model's performance during training and prevent overfitting, which occurs when the model fits too closely to the training data and performs poorly on new, unseen data. Thus, we split the initial 10,000 maps into training, validation and the test dataset.

By doing the splitting, and test sets, we can ensure that the model is trained on a diverse set of inputs and can make accurate predictions for unseen data. This is an important step in developing a reliable and accurate model for predicting $\fnllocal$ values from CMB maps. The dataset of 10,000 maps is split into 9,500 for training and 500 for testing.  The maps separated for the tests are not ``seen'' by the algorithm during the training and validation phases, and can be safely used to evaluate the model’s true performance. While training, 10\% of the training maps (950 realizations) were assigned for the validation dataset, and the remaining 8,550 maps were used to train the algorithm.

During training, two loss functions were considered: MSE and MAE. After assessing the model's performance, mean absolute error was selected as the preferred loss function, yielding slightly better results. Extensive hyperparameter tuning was performed to optimize the model, including batch size ($256$) and learning rate ($0.001$). Early stopping and model checkpoints were employed to prevent overfitting and save the best-performing weights. The training process was initially set to run for 1,000 epochs, but concluded at 909 epochs when no significant improvement in validation loss was observed. The weights from the epoch with the best performance were saved for subsequent evaluation on the test dataset.

A batch size denotes the number of samples processed simultaneously during a single training iteration in a machine-learning model. Distinct from an epoch, which encompasses a complete pass through the entire dataset, the training dataset is partitioned into smaller batches, with each batch fed into the neural network during training. For instance, if we have 1000 training samples and a batch size of 100, then 100 images are presented to the network in each iteration until all 1000 images are processed, completing one epoch. Batch size is a critical hyperparameter influencing both performance and efficiency. Smaller batch sizes may lead to slower convergence and less stable training, while larger sizes can accelerate training but may demand more memory. The choice of batch size is often determined by available computing resources.

Learning Rate is a hyperparameter that determines the step size of the parameter updates during the optimization process. It determines how much the weights of the model should be adjusted in each iteration of the optimization algorithm. The learning rate is a crucial hyperparameter in optimization algorithms, determining the size of parameter updates during training. It influences the speed of convergence and the risk of overshooting. A high learning rate can accelerate convergence but may cause instability, while a low rate may slow convergence but improve stability. Selecting the optimal learning rate depends on factors like model complexity, dataset size, loss function shape, and optimization algorithm.

This meticulous training approach ensures that the model is optimized and capable of accurately estimating $f_{\rm NL}$ on previously unseen data. Here, unseen data refers to the data that is not used in training the CNN like test dataset. 

\section{Results}
\label{sec:results}
In this section, we present the results obtained from our trained model. 
\subsection{Training and Validation loss curve}

One important factor that indicates a well-trained model is the behavior of the training and validation loss curves throughout the learning process. These curves provide valuable insights into the model's performance, including close scrutiny of either underfitting or overfitting behaviors that require further adjustments.

\begin{figure}
    \includegraphics[width=\columnwidth]{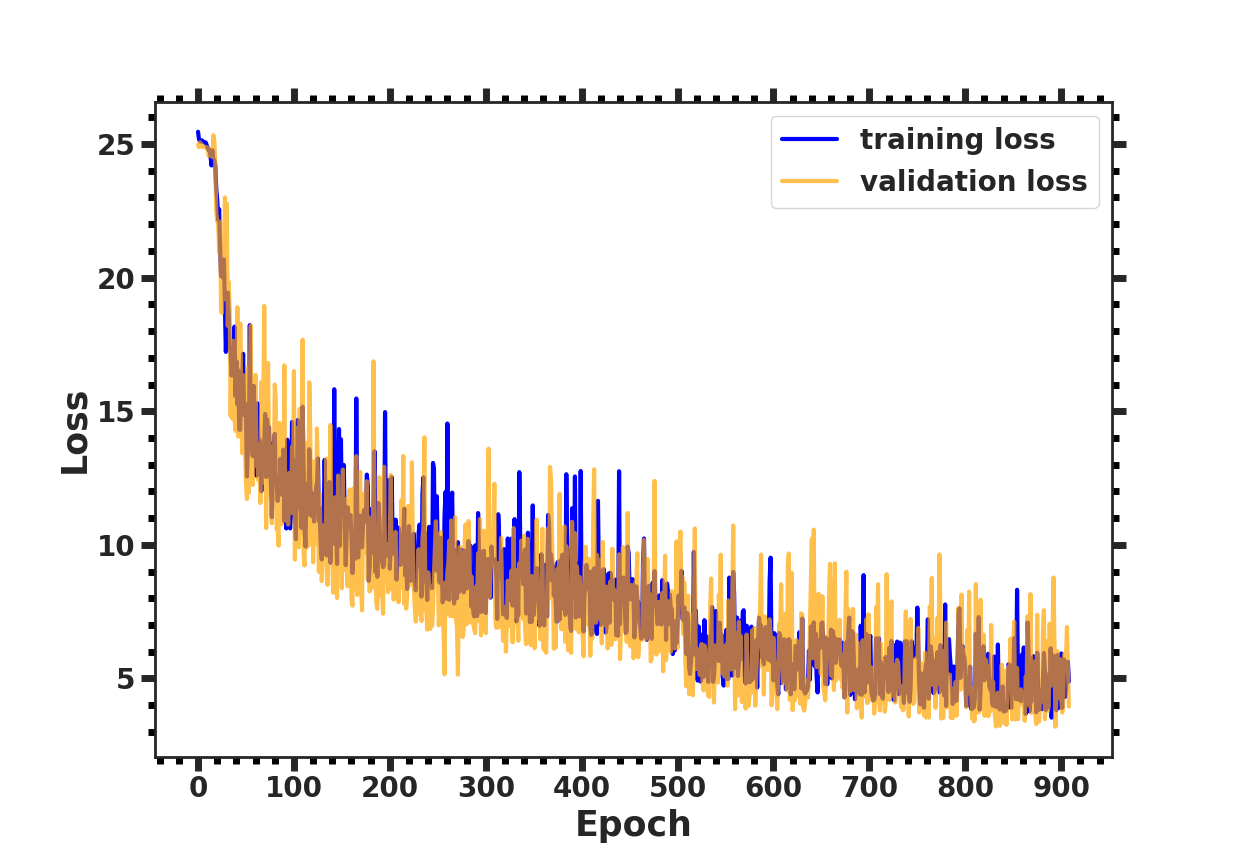}
    \caption{The image shows the training loss and validation loss for $9500$  number of maps used for training.}  
    \label{fig:Training and Validation loss}
\end{figure}

In our study, we utilized the MAE as the chosen loss function. Initially, both the training and validation loss started at values above 25. As the model began training, the loss function gradually decreased, following a smooth curve as illustrated in Fig.~\ref{fig:Training and Validation loss}. Over the course of 909 epochs, the validation loss and training loss exhibited a steady decrease until the early stopping mechanism was triggered. The initial epoch was set to 1000, and the best validation loss of 3.2 was achieved at epoch 895. Importantly, the loss curves demonstrate a desirable pattern, showing that our model is not suffering from overfitting or underfitting. In machine learning, an epoch signifies a complete iteration through the entire dataset during the training phase. Put simply, it involves presenting all the training examples to the model once. For instance, if a training dataset consists of 1000 maps and a Convolutional Neural Network (CNN) is utilized, one epoch entails feeding all 1000 images into the model for training. Following each epoch, the model's parameters are adjusted based on its performance on the dataset. This iterative process continues until the model converges on a satisfactory set of parameters that yield desirable results on the validation set. Consequently, repeating this process one thousand times constitutes 1000 epochs.

The observed behavior of the loss curves provides evidence of the effectiveness of our training process. The gradual decrease in loss signifies that the model is learning and adapting to the data, while the absence of significant fluctuations suggests stability in the training procedure. The achieved validation loss of 3.2 demonstrates the model's ability to generalize well to the unseen data. These results highlight the successful training of our model, indicating its potential for accurate predictions and reliable performance on the unseen data. 

\subsection{Model Evaluation}
\subsubsection{Estimation on synthetic data}
In the analysis, the performance of the trained model is evaluated by estimating the input $f_{\rm NL}$ values on both the training dataset and the unseen test dataset. The model checkpoint saves the weights of the epoch with the least validation loss for further analysis.

\begin{figure}
    \includegraphics[width=\columnwidth]{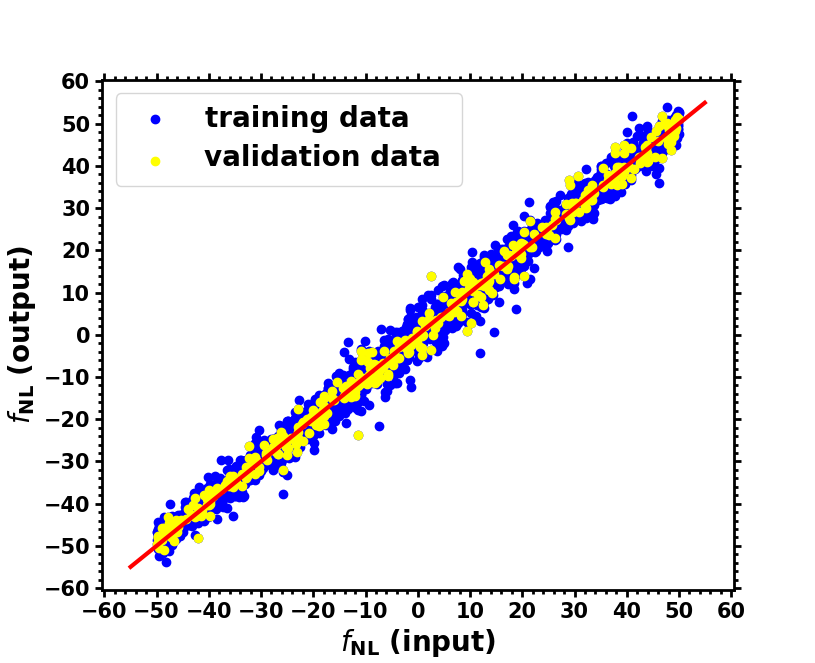}
    \caption{Scatter plot showing the comparison of estimated $f_{\rm NL}$ values by a CNN model on training and test dataset, with the solid line indicating the ideal case where estimated values would match the true values. The blue dots are training data $(n = 8550)$ and yellow dots $(n = 950)$ are validation data, the deviation from the ideal line in red indicates the level of error in the model's predictions.}
    \label{fig:fnlinout}
\end{figure}

To assess the model's performance, we plot the output $\fnllocal$ values against the input values in Fig.~\ref{fig:fnlinout}. The blue dots represent the estimation of $f_{\rm NL}$ on the training data consisting of $8,550$ $\fnllocal$ values, while the yellow dots represent the estimation on the validation data containing $950$ $\fnllocal$ split during the training. The $y$-axis corresponds to the output $f_{\rm NL}$ values predicted by the model, while the red line represents the ground truth where $y=x$.

In Fig.~\ref{fig:fnlinout}, we can see that the $\fnllocal$ outputs against inputs graph exhibits a strong correlation, demonstrating a reasonable level of accuracy. 

\subsubsection{Tests of unbiasedness}
We further test the unbiasedness of the test data for the $n=500$ output $\fnllocal$ against input values. We utilise the minimal $\chi^{2}$-fitting method, which is a straightforward way to find the best-fitting line through a series of scattering points. Considering a straight line with slope $a$ and interception $b$, we model the linear regression as $y=ax+b$, then we formulate the $\chi^{2}$ as follows
\begin{eqnarray}
\chi^{2}=\sum^{n=500}_{i=1}\left(f^{\rm out}_{\rm NL}-y(f^{\rm in}_{\rm NL}) \right)^{2}.
\end{eqnarray}

\begin{figure}
    \includegraphics[width=\columnwidth]{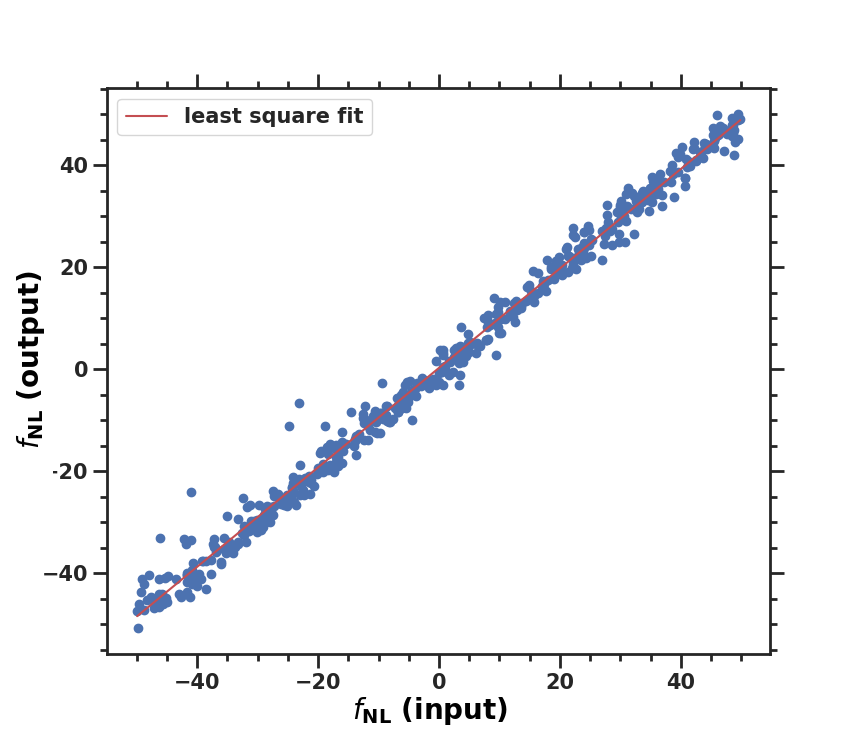}
    \caption{This graph shows the relationship between the input $f_{\rm NL}$ and the predicted $f_{\rm NL}$ values from a machine learning model. The blue dots $(n=500)$ represent the input $f_{\rm NL}$ values, while the red line represents the least square fit of the input $f_{\rm NL}$ and predicted $f_{\rm NL}$. The intercept of the line is $0.274$, and the slope is $0.975$, indicating a strong positive correlation between the two variables.}
    \label{fig:leastsq}
\end{figure}

\begin{figure}
    \centering
    \includegraphics[width=\columnwidth]{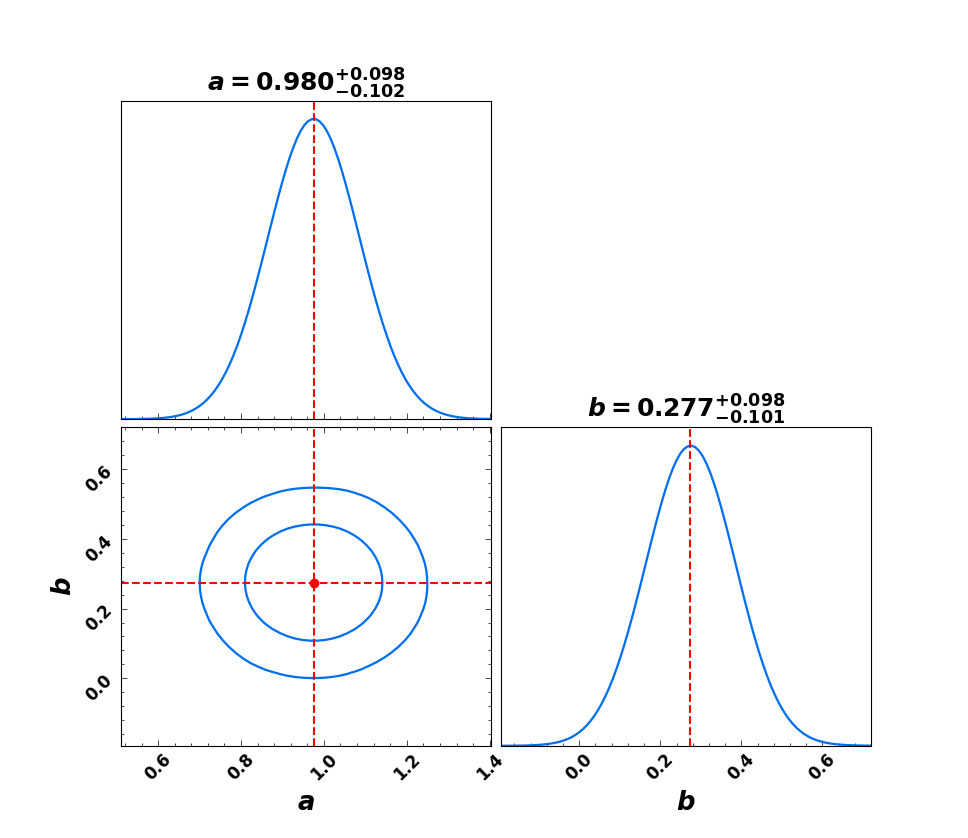}
    \caption{Constraints on slope parameter $a$ and interception $b$ by using the test data. The two 2-D contours are the $68\%$ and $95\%$ confidence level of the parameters $(a,b)$, and the diagonal plots are the marginalised distribution of the parameters.}
    \label{fig:confidence ellipse}
\end{figure}

    
Then we run it on a regular 2-D grid of parameter space $(a,b)$, with flat prior ranges as $a=(-10,10)$, $b=(-10,10)$ and each side of sampling 10,000. We show the results of the constraint in Fig.~\ref{fig:leastsq}. One can see that the best-fitting slope parameter $a$ and interception $b$ are
\begin{eqnarray}
a=0.980^{+0.098}_{-0.102}, \quad b=0.277^{+0.098}_{-0.101},    
\end{eqnarray}
which recovers the complete unbiasedness $(a,b)=(1,0)$ in $2\sigma$ C.L. One can also see the 2-D joint constraint in the corner plot in Fig.~\ref{fig:confidence ellipse}, which shows that the unbiased linear regression $(a,b)=(1,0)$ locates within the $95\%$ C.L. contour. The $R^{2}$ score for the test data is 0.96, indicating a high level of accuracy in the estimation provided by the CNN model.

\section{Conclusion}
\label{sec:conclusion}

This paper developed a deep learning model that could accurately identify the non-Gaussianity parameter ($f_{\rm NL}$), directly from cosmic microwave background (CMB) maps. To achieve this, we trained a convolutional neural network (CNN) on a synthetic CMB map dataset that contained maps with different values of $f_{\rm NL}$. The CNN model consisted of several layers, including convolutional, pooling, and fully connected layers, which were optimized using the Adam optimizer and the binary MAE (mean absolute error) loss function. Adam optimizer (Adaptive Moments Estimation) is one of the standard optimization methods used in machine learning techniques, which is basically an optimization method to minimize the loss function during the training of neural networks and subsequently to update the learning rate. There are many ways to minimize a loss function, like Gradient Descent (GD), SGD (Stochastic Gradient Descent), Root Mean Square Propogation (RMSProp), etc. Each of these method has its advantages and disadvantages relative to the problem in task. Adam optimizer combines the best of SGD and RMSProp which helps in minimizing the loss function with little computation and memory.

During training, we set the value of batch size to 256 and a learning rate to 0.001, which were selected based on their impact on the model's performance. We also employed various early stopping, to prevent overfitting to the training data.

The plot of loss function versus epoch for both training data and validation data is a smooth curve which tells us  the model is converging and not overfitting to the training data. In our case, the increase in validation loss is minimal  and does not affect the overall performance of the model. This suggests that the model has generalized well to new data and is not overfitting to the training set. We can see that the loss for both the training and validation data decreases gradually during the initial epochs, indicating that the model is learning quickly and efficiently. The fact that the loss for both datasets converges to around a similar value also suggests that the model has learned to distinguish between maps with different values of $f_{\rm NL}$ on both the training and validation sets. The smooth curve of the loss function versus epoch provides further evidence that our CNN model is well-designed and trained,

To evaluate the performance of our CNN model, we used two metrics: prediction accuracy on both the input map used during training and on a separate test map that was not seen during training. We found that the model can achieve high accuracy on both the training and test datasets, indicating that it has learned to distinguish between maps with different values of $f_{\rm NL}$.

Overall, our results suggest that a well-designed and trained CNN model can be effective in identifying $f_{\rm NL}$ from synthetic CMB map images directly. However, we acknowledge that further testing with more diverse and realistic data with varying cosmological parameters are necessary to fully assess the model's performance and generalizability. Nonetheless, our results are promising and demonstrate the potential of deep learning techniques for cosmological parameter estimation from the CMB map directly.

\section*{DATA AVAILABILITY}
The pipeline that generates the training and testing data was developed by our research group, and can be shared upon reasonable request to the corresponding author. The simulated raw non-Gaussian maps used in the analysis is public and downloadable at \url{http://dc.zah.uni-heidelberg.de/elsnersim/q/s/fixed} by properly crediting~\citet{Elsner2009}. 

\section*{Acknowledgements} 
This research is funded by the research program ``New Insights into Astrophysics and Cosmology with Theoretical Models Confronting Observational Data'' of the National Institute for Theoretical and Computational Sciences of South Africa. All computations were carried out using the computational cluster resources at the Centre for High-Performance Computing, Cape Town, South Africa. Y.Z.M. acknowledges the support of the National Research Foundation with grant No.~150580. We also acknowledge the usage of non-Gaussianity map simulations provided by~\citet{Elsner2009}. C.G.N acknowledges Dr. Guo-Jian Wang and Dr. Cheng Cheng for useful discussions.

\bibliographystyle{mnras} 
\bibliography{ML_nG_ref} 
\bsp	
\label{lastpage}
\end{document}